\documentclass[11pt]{article}
\pdfoutput=1 % if your are submitting a pdflatex (i.e. if you have
\usepackage{jcappub} % for details on the use of the package, please
\usepackage[T1]{fontenc} % if needed

\usepackage{amsmath}
\usepackage{amsfonts}
\usepackage{amssymb}
\usepackage{graphicx,color}
\usepackage{subfigure}
\usepackage{hyperref}
\usepackage[usenames,dvipsnames,svgnames,table]{xcolor}
\usepackage{enumerate}

\usepackage[left=2cm,top=2.5cm,right=2cm,bottom=2cm]{geometry}
\def \be  {\begin{equation}}
\def \ee  {\end{equation}}
\def \ba  {\begin{eqnarray}}
\def \ea  {\end{eqnarray}}

\def \Mpl {M_{\rm Pl}}

\title{Canonical single field slow-roll inflation with a non-monotonic tensor-to-scalar ratio} 
\author[a,b,\star]{Gabriel Germ\'an,}
\author[c,d]{Alfredo Herrera--Aguilar} 
\author[b]{Juan Carlos Hidalgo}
\author[e]{and Roberto A. Sussman}
\affiliation[a]{Rudolf Peierls Centre for Theoretical Physics, University of Oxford, \\1 Keble Road, Oxford, OX1 3NP, UK}
\affiliation[b]{Instituto de Ciencias F\'isicas, Universidad Nacional Aut\'onoma de M\'exico, \\
Apdo. Postal 48-3, 62251 Cuernavaca, Morelos, M\'{e}xico.}
\affiliation[c]{Instituto de F\'{\i}sica, Benem\'erita Universidad Aut\'onoma de Puebla, \\ Apdo. postal J-48, CP 72570, Puebla, Pue., M\'exico}
\affiliation[d]{Instituto de F\'{\i}sica y Matem\'{a}ticas,
Universidad Michoacana de San Nicol\'as de Hidalgo,\\
Edificio C--3, Ciudad Universitaria, CP 58040, Morelia, Michoac\'{a}n, M\'{e}xico.}
\affiliation[e]{Instituto de Ciencias Nucleares,
{{Universidad Nacional Aut\'onoma de M\'exico, }}
{{Apdo. Postal 70-543, 04510 M\'exico D. F., M\'{e}xico.}}}
\affiliation[\star]{On sabbatical leave from
Instituto de Ciencias F\'isicas, Universidad Nacional Aut\'onoma de M\'exico. Corresponding author: gabriel@fis.unam.mx.}
\abstract{We take a pragmatic, model independent approach to single field slow-roll canonical inflation by imposing conditions,
not on the potential, but on the slow-roll parameter $\epsilon(\phi)$ and its derivatives $\epsilon^{\prime }(\phi)$ and $\epsilon^{\prime\prime }(\phi)$, thereby extracting general 
conditions on the tensor-to-scalar ratio $r$ and the running $n_{sk}$ at $\phi_{H}$  where the perturbations are produced, some $50$ $-$ $60$ $e$-folds before the end of 
inflation. We find quite generally that for models where $\epsilon(\phi)$ develops a maximum, a relatively large $r$ is most likely accompanied by a positive running while a negligible tensor-to-scalar ratio implies negative running. The definitive answer, however, is given in terms of the slow-roll parameter $\xi_2(\phi)$. 
To accommodate a large tensor-to-scalar ratio that meets the limiting values allowed by the Planck data, we study a non-monotonic $\epsilon(\phi)$ decreasing during most part of inflation. Since at $\phi_{H}$ the slow-roll parameter $\epsilon(\phi)$ is increasing, we thus require that $\epsilon(\phi)$ develops a maximum for $\phi > \phi_{H}$ after which 
$\epsilon(\phi)$ decrease to small values where most $e$-folds are produced. The end of inflation might occur trough a hybrid mechanism and a 
small field excursion $\Delta\phi_e\equiv |\phi_H-\phi_e |$ is obtained with a sufficiently thin profile for $\epsilon(\phi)$ which, however, should not conflict with the second 
slow-roll parameter $\eta(\phi)$. As a consequence of this analysis we find bounds for $\Delta \phi_e$, $r_H$ and for the scalar spectral index $n_{sH}$. Finally we provide examples where these considerations are explicitly realised.}
\begin{document}
\maketitle
\flushbottom

\section{Introduction} \label{Intro} 

\noindent 

\label{Introsec}

Inflation \cite{Guth:1981}-\cite{Albrecht:1982}, \cite{Lyth:1998xn} has proved to be very useful in explaining not only the homogeneity 
of the universe on very large scales but also in providing a theory of structure formation. Typically, slow-roll models of inflation 
are specified by a formula for the potential which in the single field case generically predicts Gaussian, adiabatic and nearly scale-invariant primordial fluctuations.
Here, instead of testing the observables for a specific potential we study general characteristics of the inflationary paradigm  by looking at properties of the slow-roll parameter $\epsilon(\phi)$ and its 
derivatives with respect to $\phi$ denoted by $\epsilon(\phi)^{\prime }$ and $\epsilon(\phi)^{\prime\prime }$. 
In what follows we concentrate on 
single field slow-roll canonical inflation (for noncanonical kinetic studies see \cite{Hu:2011vr}).

Observable scales of primordial perturbations were produced some $50$ $-$ $60$ $e$-folds before the end of inflation. We denote quantities at this scale with the subscript $H$. 
When the tensor-to-scalar ratio $r$ is large, 
$r_H\approx 0.12$ (taking the upper limit of the Planck \cite{Ade:2015lrj} or Planck-Keck-BICEP2 \cite{Ade:2015tva} analysis) a slightly modified Lyth bound \cite{Lyth:1997} implies 
a relatively large range of the inflaton excursion  $\Delta\phi_8\equiv |\phi_H-\phi_8|$ for the observable cosmological scales $\Delta N_8 \approx 8$ and for an increasing $\epsilon(\phi)$ during the first few $e$-folds of observable inflation,  $\epsilon(\phi)\ge r_{H}/16$, thus
$\Delta\phi_8 \approx \Delta N_8\sqrt{2\epsilon(\phi)}\Mpl \geq 8 \sqrt{r_{H}/8} \Mpl \approx 0.98 \Mpl$. Here $\Mpl$ is the reduced Planck mass $\Mpl=2.44\times 10^{18} \, \mathrm{GeV}$ which we set $\Mpl=1$ in what follows.
In the Boubekeur-Lyth bound 
\cite{Boubekeur:2005zm}, a stronger result follows when $\epsilon(\phi)$ does not decrease during inflation. 
To have a small $\Delta\phi_e\equiv |\phi_H-\phi_e|$ with a relatively large tensor-to-scalar ratio it seems that 
we are invited to consider a decreasing $\epsilon(\phi)$ during most part of inflation with a large number of $e$-folds generated 
not around $\phi_{H}$ but close to the end of inflation at $\phi_e$ (for work in this direction see e.g., 
\cite{BenDayan:2009kv}-\cite{Chatterjee:2014hna}). The present paper explores the possibility of a maximum in the evolution of $\epsilon(\phi)$\footnote{This idea has been suggested by \cite{Ross:2009hg} where a non-monotonic tensor-to-scalar ratio with a maximum occurs in a very natural way.} with particular attention to the consequent values for $\Delta \phi_e$ and $r_H$. It is clear for instance that $\epsilon(\phi)$ should not decrease too much because 
then the small-scale power spectrum becomes so large that primordial black holes are overproduced \cite{Kohri:2007qn}, \cite{Vazquez:2014uca}.
Moreover, the end of inflation in this case should be achieved not by the inflaton-field itself, but by some other mechanism e.g, an hybrid field, although all of inflation is driven by a single field.

 This article is organised as follows: In Section 2 we discuss general consequences of a non-monotonic tensor-to-scalar ratio during observable inflation, in particular for the running $n_{sk}$ defined by Eq.~(\ref{Slownsk}) below. Section 3 contains a discussion of bounds for $\Delta\phi_e$ and $r_H$ while in Section 4 we provide two examples of well motivated models one with a monotonic and another with a non-monotonic tensor-to-scalar ratio. Finally in Section 5 one can find a summary of our results and concluding remarks.
%%%%%%%%%%%%%%%%%%%%%%%%%%%%%%%%%%%%%%%%%%%%%
%%%%%%%%%%%%%%%%%%%%%%%%%%%%%%%%%%%%%%%%%%%%%
\section{The scalar spectral index and the running}  \label{running}

\begin{figure}[!t]
\begin{centering}
\includegraphics[scale=0.6]{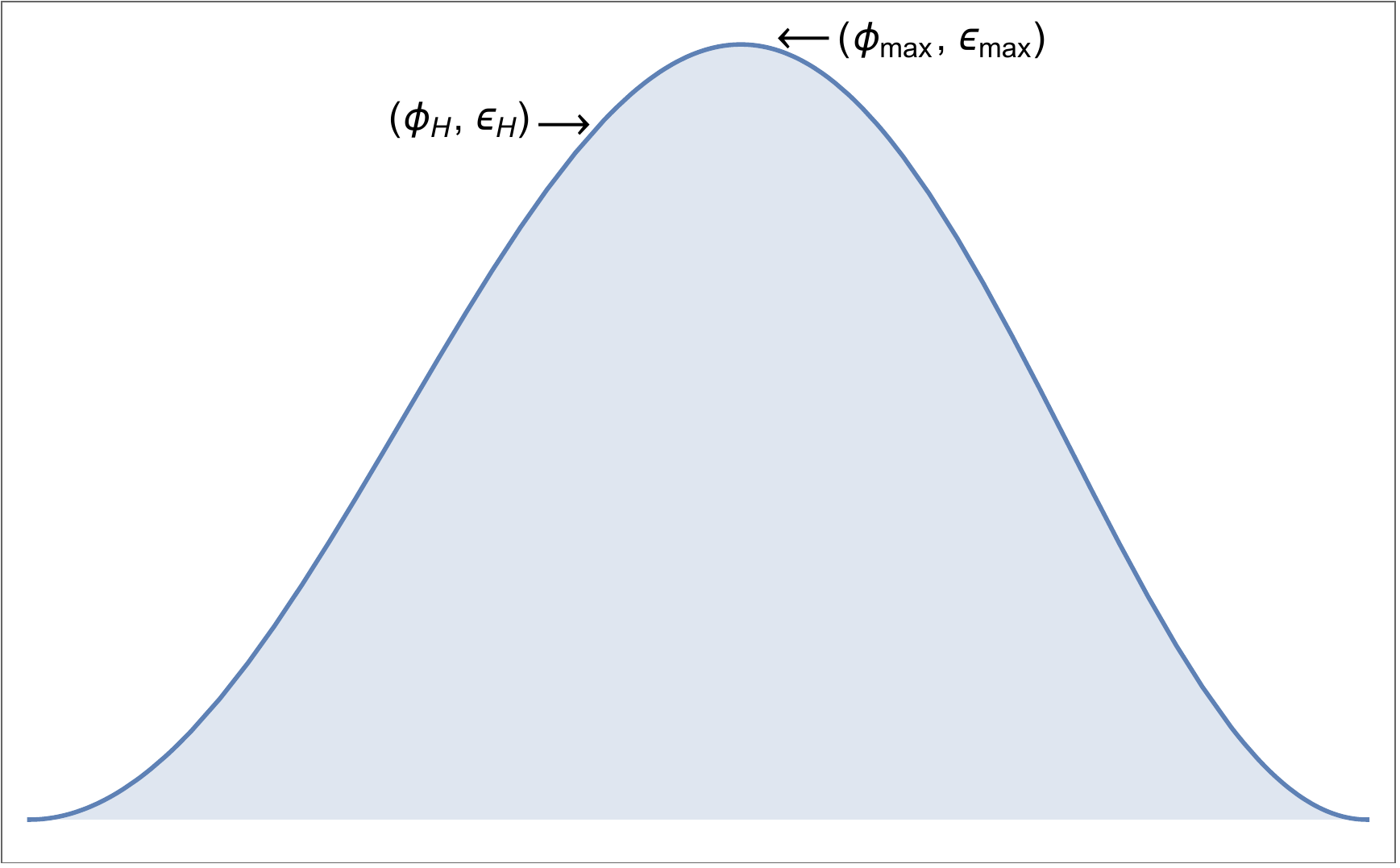}
\caption{\small Plot of the slow-roll parameter $\epsilon(\phi)$ showing a maximum for $\phi = \phi_{max}$, close to $\phi_{H} $. The maximum is required because, 
for observable inflation, $\epsilon(\phi)$ is increasing at $\phi_H<\phi_{max}$. At least 8 $e$-folds of inflation should occur in the interval  $\phi_H<\phi<\phi_{max}$. 
For $\phi>\phi_{max}$, $\epsilon(\phi)$ can decrease practically generating all of inflation. The maximum of $\epsilon(\phi)$ can not occur at  $\phi_{max} = \phi_H$ because
the value of the tensor-to-scalar ratio $r$ would violate observational bounds. The end of inflation for vanishing $\epsilon(\phi_e)$ is triggered by a hybrid field and a small 
$\Delta\phi_e$ is obtained when the profile depicted for $\epsilon(\phi)$ is sufficiently \emph{thin} which, however, should not conflict with the curvature of the potential, controlled by the slow-roll parameter $\eta$. }
\label{f1}
\end{centering}
\end{figure}

Here we study consequences of a non-monotonic tensor-to-scalar ratio for the spectral index and for the running. The slow-roll parameters \cite{Liddle:2000cg} which involve the potential and its derivatives\footnote{From now on we drop the $\phi$-dependence but keep the $H$-subindex label to emphasize that quantities are evaluated at the scale $H$.} are defined by
\begin{equation}
\epsilon \equiv \frac{1}{2}\left( \frac{V^{\prime }}{V }\right) ^{2},\quad
\eta \equiv \frac{V^{\prime \prime }}{V}, \quad
\xi_2 \equiv \frac{V^{\prime }V^{\prime \prime \prime }}{V^{2}},
\label{Slowpara}
\end{equation}%
where primes denote derivatives with respect to $\phi$.  In the slow-roll approximation the scalar spectral index and the running are given in terms of the 
usual slow-roll parameters \cite{Liddle:2000cg} as follows 
\begin{eqnarray}
n_{s} &=&1+2\eta -6\epsilon ,  \label{Slowns} \\
n_{sk} &=&\frac{d n_{\mathrm{s}}}{d \ln k}=16\epsilon \eta -24\epsilon ^{2}-2\xi_2. \label{Slownsk} 
\end{eqnarray}
There is evidence that the power-spectrum over the range of observable scales is decreasing in amplitude as the scales decrease which means that, while this range of scales were leaving the horizon during inflation, $\epsilon$ was increasing. For a non-monotonic $\epsilon$, a simple but interesting possibility  is  $\epsilon$ developing a maximum at $\phi_{max}$ during inflation (see Fig.~\ref{f1}). First we realize that $\phi_H$ cannot be located at $\phi_{max}$ because this would be inconsistent with the Planck data \cite{Ade:2015lrj}: $\delta _{ns}> 0.0307$ together with the result $r = 8 \delta _{ns}$ at the maximum, would imply  $r > 0.246$. Thus, $\phi_H < \phi_{max}$ and $\epsilon_H < \epsilon_{max}$, which is consistent with an increasing $\epsilon$ during observable inflation. 
The derivative of the slow-roll parameter $\epsilon$ at its maximum is
\begin{equation} 
\epsilon^{\prime }\equiv\frac{d \epsilon}{d\phi} =\frac{V^{\prime }} {V }\left(\eta-2\epsilon\right) = 0, \quad  \Longleftrightarrow \quad 
\eta_{max} = 2\epsilon_{max} .
\label{epsmaxeq}
\end{equation}
As usual, the second derivative of 
$\epsilon$ at $\eta = 2\epsilon$ characterises this critical point
\begin{equation} 
\epsilon^{\prime\prime }\Bigl|_{\eta=2\epsilon}\equiv\frac{d^2 \epsilon}{d\phi^2} \Bigl|_{\eta=2\epsilon}=\left[\xi_2-2\eta\epsilon-4\epsilon\left(\eta-2\epsilon\right) +\left(\eta-2\epsilon\right)^2\right]
\Bigl|_{\eta=2\epsilon} =\xi_2-4\epsilon^2,
\label{segundaderieps1}
\end{equation}
which for a maximum implies that $\xi_{2\,max} < 4\epsilon_{max	}^2$.                 

Let us concentrate for the moment in the expression where $ V^{'}<0$ with the potential $V$ a monotonically decreasing 
function of $\phi$ during inflation. In this case $\phi$ is evolving away from the origin, thus the derivative of 
$\epsilon$ is
\begin{equation}
\epsilon^{\prime } = -\sqrt{2\epsilon}\left( \eta -2\epsilon\right) .
\label{epsprima2}
\end{equation}%
The case $ V^{'}>0$ would correspond to $\phi$ evolving towards the origin and can be analysed in a similar way. 
In what follows we parameterise the deviation from the Harrison-Zeldovich spectrum with  $\delta _{ns}$ 
defined by $\delta_{ns}\equiv 1-n_{s}$. We note that Eq.~(\ref{epsprima2}) together with Eq.~(\ref{Slowns}) can be thus written as
\begin{equation}
\epsilon^{\prime } = \frac{1}{2}\sqrt{2\epsilon}\left( \delta _{ns} -2\epsilon\right).
\label{epsprima3}
\end{equation}%

Requiring a non-decreasing $\epsilon$ during observable inflation means that there should be at least 8 $e$-folds of inflation between $\phi_{\mathrm{H}}$ and 
the maximum of $\epsilon$ at $\phi_{max}$. This implies $\epsilon^{\prime }>0$ at least during this range, or 
\begin{equation}
\delta _{ns} > 2\epsilon > 2\epsilon_{H}>0, \
\label{deltaeps}
\end{equation}% 
which means that during this window of observable inflation the spectral index is bounded by
\begin{equation}
n_{s}  <  1-2\epsilon_{H}=1-r_H/8 .
\label{nsbound1}
\end{equation}%
In the interval $\phi_H < \phi < \phi_{max}$, $\epsilon$ grows from $\epsilon_H$ to $\epsilon_{max}$ but its derivative decreases vanishing at $\phi_{max}$ thus $\epsilon_H < \epsilon$ and $\epsilon'_H > \epsilon' .$ From Eq.~(\ref{epsprima3}) we get $1-n_{s}-2\epsilon <\frac{\sqrt{\epsilon_H}}{\sqrt{\epsilon}}(1-n_{sH}-2\epsilon_H)<1-n_{sH}-2\epsilon_H,$ or 
\begin{equation}
 n_{s_H}  < n_{s}  +2(\epsilon-\epsilon_H),
\label{nsbound2}
\end{equation}%
where $\epsilon-\epsilon_H>0$. Thus, although $\epsilon$ is larger than $\epsilon_H$, $n_s$ is not constrained to be smaller than $n_{sH}$ due to the contribution of the $\eta$-term present in $n_s$ ($\eta_H$-term present in $n_{sH}$).

To find possible consequences for the running let us now consider the second derivative of $\epsilon$. Together with Eq.~(\ref{Slowns}), we have
\begin{eqnarray} 
\epsilon^{\prime\prime } &=&\xi_2-2\eta\epsilon-4\epsilon\left(\eta-2\epsilon\right) +\left(\eta-2\epsilon\right)^2 = 
-9 \epsilon^2 + 2\epsilon\delta _{ns}+\frac{1}{4}\delta _{ns}^2+\xi_2 \notag \\
&=& \frac{1}{8}\left(-\frac{9}{32}r^2+r\delta _{ns}+2\delta _{ns}^2\right)+\xi_2.
\label{segundaderieps3}
\end{eqnarray}
In a similar way, we write the running as 
\begin{equation} 
n_{sk} =16\epsilon \eta -24\epsilon ^{2}-2\xi_2 = 24\epsilon ^{2}-8\epsilon\delta _{ns}-2\xi_2 =\frac{r}{2}\left(\frac{3}{16}r-\delta _{ns}\right)-2\xi_2.
\label{Slownsk2}
\end{equation}
According to the Planck data (last column of Table 4 of \cite{Ade:2015lrj}), at $\phi_H$, $r_H<0.15$, $n_{sH}=0.9644\pm 0.0049$, and $n_{skH}=-0.0085\pm 0.0076$.  Thus, $\frac{3}{16}r_H-\delta _{nsH}<0$.
The running $n_{skH}$ will be negative if and only if
\begin{equation} 
\xi_{2H} > \frac{r_H}{4}\left(\frac{3}{16}r_H-\delta _{nsH}\right)>-5.5 \times 10^{-4}.
\label{xi2}
\end{equation}
In particular, a positive $\xi_{2H}$ will always give a negative running and from Eq.~(\ref{segundaderieps3}), since $-\frac{9}{32}r_H^2+r_H\delta _{nsH}+2\delta _{nsH}^2>0$, a positive $\epsilon_H^{\prime\prime}$. 
Furthermore, substituting $\xi_2$ from Eq.~(\ref{segundaderieps3}) into Eq.~(\ref{Slownsk2})
\begin{equation} 
n_{sk} =-2\epsilon^{\prime\prime } +\frac{1}{128}\left(r-8\delta _{ns}\right)\left (3 r-8\delta _{ns}\right).
\label{Slownsk3}
\end{equation}
Thus, we conclude that, in general, $n_{sk}$ will be negative if and only if
\begin{equation} 
\epsilon^{\prime\prime }>\frac{1}{256}\left(r-8\delta _{ns}\right)\left (3 r-8\delta _{ns}\right).
\label{epp1}
\end{equation}

The last inequality implies mostly positive values for  $\epsilon_H^{\prime\prime }$ except in the region where 
$\delta _{nsH}/6 < \epsilon_H< \delta _{nsH}/2$, equivalently $8\, \delta _{nsH}/3 < r_H< 8\, \delta _{nsH}$ or $0.082 < r_H< 0.246$, which is still within the bound $r_H<0.15$ when running is allowed (see Table 4 of \cite{Ade:2015lrj}), but interestingly enough, not a zero value either. 

\begin{figure}[!ht]
\begin{centering}
\includegraphics[scale=0.5]{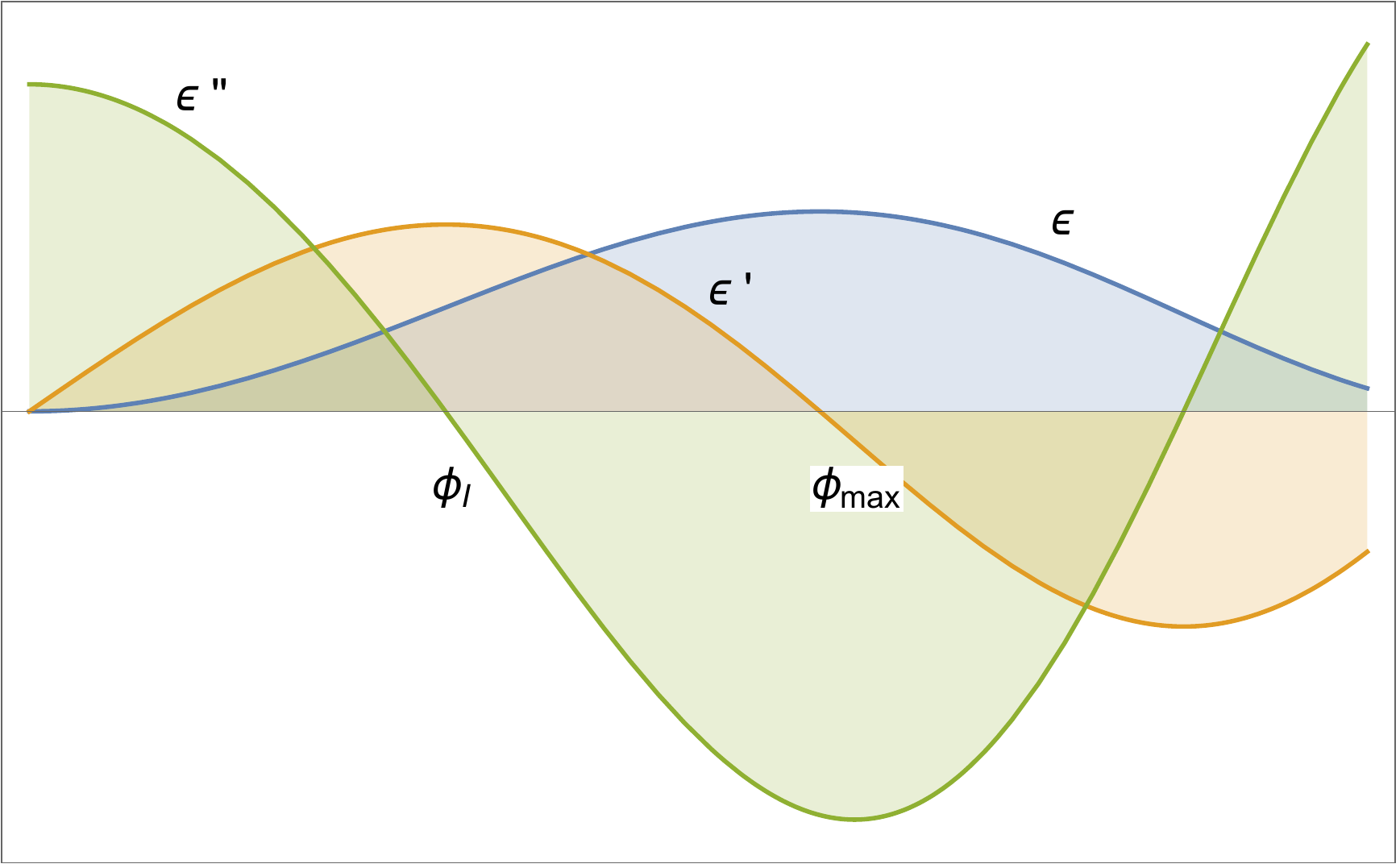}
\caption{\small Plot of the slow roll parameter $\epsilon(\phi)$ and its derivatives $\epsilon^{\prime}(\phi) $ and $\epsilon^{\prime\prime }(\phi) $ with respect to $\phi$, as functions of the field, for a model where $\epsilon(\phi)$ is a non-monotonic function of $\phi$. The labels $\phi_{max}$ and $\phi_I$ refer to the values of $\phi$ for which the maximum and the inflection points of $\epsilon(\phi)$ occur, 
respectively.} 
\label{f2}
\end{centering}
\end{figure}

When approaching the maximum of $\epsilon$ its derivative $\epsilon^{\prime }$ tends to zero with negative $\epsilon^{\prime\prime }$. A positive value for $\epsilon^{\prime\prime }$ can occur only if in addition $\epsilon(\phi)$ has an inflection point at $\phi_I<\phi_{max}$, such that $\epsilon^{\prime }(\phi_I)$ is a maximum and $\epsilon^{\prime\prime }(\phi_I)=0$, (see Fig.~\ref{f2})\,.  In this situation $\phi_H$ can be smaller or larger than $\phi_I$.
Thus, for models where $\epsilon$ presents a maximum with an inflection point as in Fig.~\ref{f2}\,,  a relatively large $r_H$ is most likely accompanied by a negative $\epsilon^{\prime\prime}$ with positive running, while a negligible tensor-to-scalar ratio implies positive $\epsilon^{\prime\prime}$ and negative running. The definitive answer, however, is to be found in the inequality given by Eq.~(\ref{xi2}) which imposes precise constraints on the parameters of an specific model.
%%%%%%%%%%%%%%%%%%%%%%%%%%%%%%%%%%%%%%%%%%%%%
%%%%%%%%%%%%%%%%%%%%%%%%%%%%%%%%%%%%%%%%%%%%%
\section{Bounds on $\Delta\phi$ and $r$ }  \label{Increasingdsec}

While observable scales were leaving the horizon during inflation, $\epsilon$ was an increasing function of $\phi$ and the Lyth bound is regarded as an inevitable consequence.
However, if for subsequent $\phi$, we study a {\it decreasing} 
$\epsilon$, then $\epsilon$ has to go through a maximum at $\phi_{max} > \phi _{H}$ before starting to decrease as in Fig.~\ref{f1}. 

The value $\phi _{H}$ at which $n_{sH} \approx 0.968$ lies to the left of $\phi_{max}$ so that 
$\epsilon^{\prime }_{H}$ is positive although small. We observe that an $\epsilon$  with the behaviour shown in 
Fig.~\ref{f1} has the potential to generate relatively large values of $r_{H}$ while sufficient inflation is produced away 
from $\phi_{H}$.

Definig $\Delta\phi_{max}\equiv | \phi_H-\phi_{max} |$ and $\Delta N_{max}$ as the corresponding number of $e$-folds generated during the field excursion of width $\Delta \phi_{max}$ the usual expression for the number of $e$-folds gives
\begin{equation}
\Delta N_{max}=\int_{\phi_H}^{\phi_{max}} \frac{d\phi}{\sqrt{2\epsilon}} > \Delta \phi_{max} \,\, \mathrm{min}\left\{ \frac{1}{\sqrt{2\epsilon}}\right\}=  \Delta \phi_{max} \,\,  \frac{1}{\sqrt{2\, \mathrm{max}\left\{\epsilon\right\}}}=   \frac{\Delta \phi_{max}}{\sqrt{2 \epsilon_{max}}},
\label{Nmax1}
\end{equation}
where $\mathrm{min}\left\{...\right\}$ and $\mathrm{max}\left\{...\right\}$ denote the minimum/maximum numerical value of the corresponding quantity in the interval 
$\Delta\phi_{max}\equiv | \phi_H-\phi_{max} |$.
However from Eq.~(\ref{epsprima3}) we have that $ 2\epsilon_{max}=\delta_{ns, max}$ thus, Eq.~(\ref{Nmax1}) becomes
\begin{equation}
\Delta\phi_{max}  < \Delta N_{max}  \sqrt{\delta _{ns, max}}\,.
\label{Deltaboundedmax}
\end{equation}%
In a similar way 
\begin{equation}
\Delta N_{max}=\int_{\phi_H}^{\phi_{max}} \frac{d\phi}{\sqrt{2\epsilon}} < \Delta \phi_{max} \,\, \mathrm{max} \left\{\frac{1}{\sqrt{2\epsilon}}\right\}= \Delta \phi_{max} \,\, \left(\frac{1}{\sqrt{2 \, \mathrm{min}\{\epsilon\}}}\right)=  \frac{\Delta \phi_{max}}{\sqrt{2 \epsilon_H}}.
\label{Nmax2}
\end{equation}

\noindent Consequently, the (slighlty modified) Lyth bound for the interval $\Delta \phi_{max}$ reads 
\begin{equation}
\Delta N_{max}\sqrt{\frac{r_H}{8}} <  \Delta \phi_{max}.
\label{eps}
\end{equation}
Combining Eqs.~\eqref{Deltaboundedmax} and \eqref{eps} we find a bounded excursion:
\begin{equation}
\Delta N_{max} \sqrt{\frac{r_{H}}{8}} < \Delta\phi_{max}  < \Delta N_{max}  \sqrt{\delta _{ns, max}}\,.
\label{Deltabounded3}
\end{equation}%

One can also write just the first inequality as a bound for $r_H$ which, for the desired upper bound $\Delta\phi_e<1$ together with $\Delta\phi_{max} <\Delta\phi_e<1$ and $\Delta N_{max} \approx 8$ $e$-folds of observable inflation yields
\begin{equation}
r_{H} \leq 8\left(\frac{\Delta\phi_{max} }{\Delta N_{max} }\right)^2 <8\left(\frac{\Delta\phi_e}{\Delta N_{max} }\right)^2<\frac{1}{8}\approx 0.12.
\label{Deltabounded3a}
\end{equation}%

\noindent This coincides very closely with Planck's bound. Thus, allowing for an increasing $\epsilon$ during observable inflation with a subsequent decrease, one should be able to 
construct models where $\Delta\phi_e < 1$ and with $r_H$ saturating the constraints imposed by Planck or Planck-Keck-BICEP2 data.

After the maximum at $\phi_{max} > \phi _{H}$ has been reached, $\epsilon$ starts decreasing up to a point where $1/\sqrt{2\epsilon}$ has generated sufficient $e$-folds and inflation is terminated by a waterfall field. 
The profile of $\epsilon$ in Fig. \ref{f1} implies that 
close to  the end of inflation, at $\phi_{e}$, the potential becomes very flat and a hybrid mechanism should terminate inflation. It is also evident that a small $\Delta\phi_e$ is obtained with a sufficiently steep drop in the value of $\epsilon$, thus a feature in 
$\epsilon$ should not only be high with a low end value but also sharp. We would expect that a sharp feature in $\epsilon$ would conflict with the slow-roll parameter $\eta$ which accounts for the curvature of the potential. 
In order to control large values of $\eta$ after $\phi$ has reached the maximum, let us rewrite Eq.~(\ref{epsprima2}) in a 
more convenient form as
\begin{equation}
\eta=2\epsilon-\frac{\epsilon^{\prime } }{\sqrt{2\epsilon}}, \quad \quad \phi > \phi_{max}.
\label{eta}
\end{equation}%
After the maximum $\phi_{max}$, $\epsilon'$ becomes negative. Consequently, both terms $2\epsilon$ and $-\frac{\epsilon^{\prime } }{\sqrt{2\epsilon}}$ in Eq.~(\ref{eta}) are positive and thus 
$0<\eta<1$ during inflation with a decreasing $\epsilon$. The first term is negligible w.r.t. the second because we want a large $\epsilon^{\prime}$, 
thus for $\phi > \phi_{max}$,  $2\epsilon$ decreases while $-\frac{\epsilon^{\prime } }{\sqrt{2\epsilon}}$ grows large
\begin{equation}
\eta=2\epsilon-\frac{\epsilon^{\prime } }{\sqrt{2\epsilon}} \approx -\frac{\epsilon^{\prime } }{\sqrt{2\epsilon}} 
\approx \frac{\Delta\epsilon_c}{\Delta\phi_c\sqrt{2\epsilon}}.
\label{etaapprox}
\end{equation}%
where $\Delta \epsilon_c\equiv |\epsilon_{max}-\epsilon_e|$ and $\Delta\phi_c\equiv |\phi_{max}-\phi_e|$ denote quantities in the complementary range of inflation. Thus, demanding that inflation is sustained for an interval $\Delta \phi_c$ (i.e. $\eta < 1$), sets a lower bound for
\begin{equation}
\Delta\phi_c \approx \frac{\Delta\epsilon_c}{\sqrt{2\epsilon}}\, \frac{1}{\eta} > \frac{\Delta\epsilon_c}{\sqrt{2\epsilon}} > 
\frac{\epsilon_{max}-\epsilon_{e}}{\sqrt{2\epsilon_{H}}} 
\approx \frac{\epsilon_{max}}{\sqrt{2\epsilon_{H}}} \approx \left(\frac{\epsilon_{H}}{2}\right)^{\frac{1}{2}}.
\label{Deltafi}
\end{equation}%
In the equation above $\epsilon<\epsilon_{max}$, thus $1/\sqrt{\epsilon} > 1/\sqrt{\epsilon_{max}}\approx 1/\sqrt{\epsilon_{H}}$\footnote{The assumption $\epsilon_H\approx\epsilon_{max}$ is justified because we expect most $e$-folds to occur for small $\epsilon$ or, equivalently, large $1/\sqrt{2\epsilon}$. While closely after $\phi_{max}$, one could have $\epsilon_{max}> \epsilon(\phi)> \epsilon_H$, the required quick drop in $\epsilon$ after $\phi_{max}$ means that for most of the excursion $\Delta\phi_c$ the value of $\epsilon$ lies below $\epsilon_H$.}. Using the bound of Eq.~(\ref{Deltabounded3a}) we get  
$\Delta\phi_c > 0.06$, for $\phi>\phi_{max}$ and $\epsilon$ decreasing. 
%%%%%%%%%%%%%%%%%%%%%%%%%%%%%%%%%%%%%%%%%%%%%
%%%%%%%%%%%%%%%%%%%%%%%%%%%%%%%%%%%%%%%%%%%%%
\begin{figure}[h!]
\label{ej1}
 \begin{center}
   \includegraphics[ width=8cm, height=6cm]{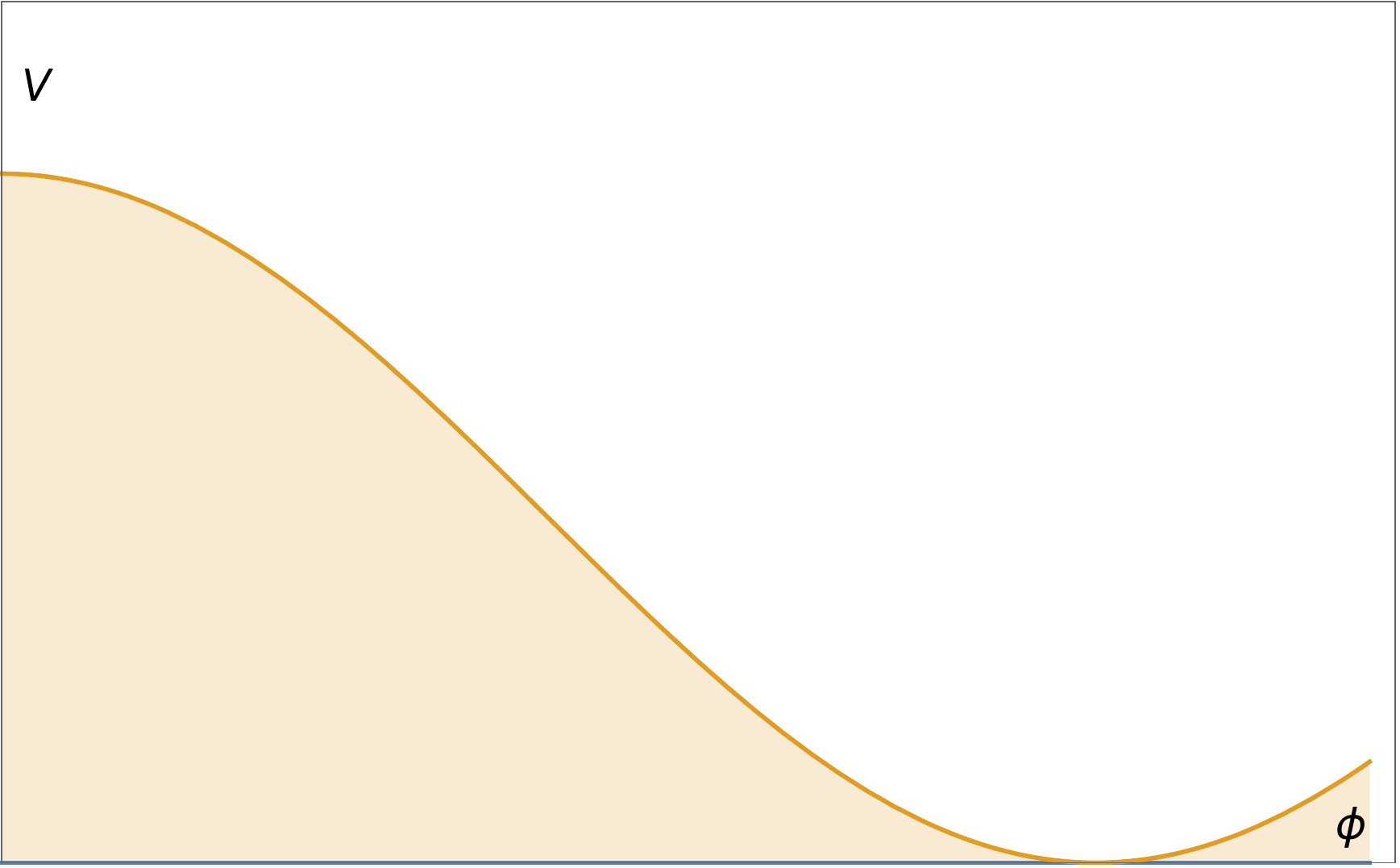}
    \includegraphics[ width=8cm, height=6cm]{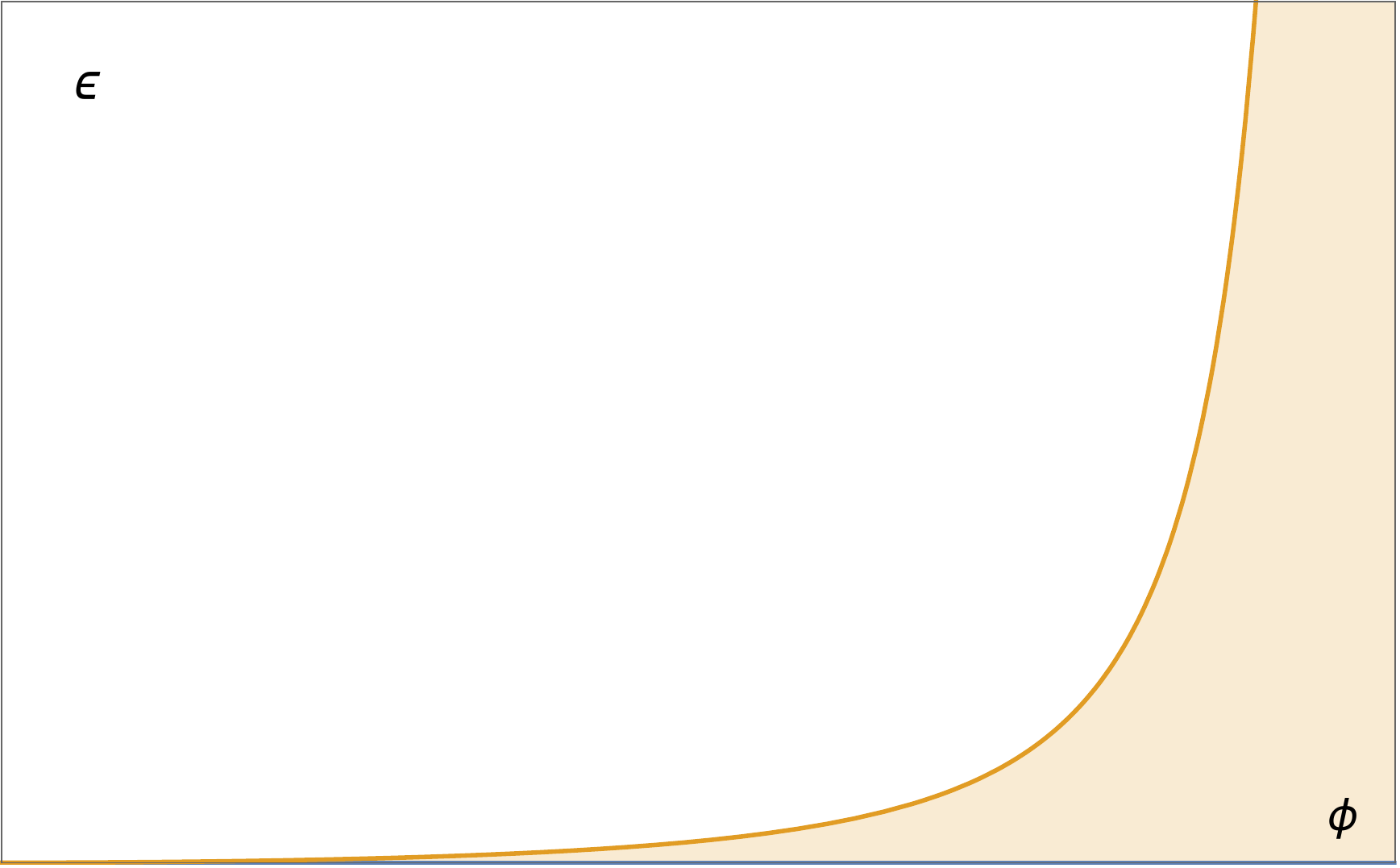}
        \caption{The potential of Natural Inflation with its corresponding tensor-to-scalar ratio (right). Because the minimum of the potential occurs with vanishing energy the (monotonic) tensor-to-scalar ratio diverges at that point. In NI inflation is terminated by the saturation of the condition $\epsilon=1$ with the inflaton also in charge of ending inflation. }
 \label{FNI}
 \end{center}
\end{figure}

\section{Examples of models with monotonic and non-monotonic tensors} \label{ex}
We illustrate some of the previous discussion with two well motivated models of inflation:  Natural Inflation (NI) \cite{Freese:1990rb}-\cite{Freese:2014nla} and Hybrid Natural Inflation (HNI) \cite{Ross:2009hg}-\cite{Ross:2016hyb}.

\subsection{A model with a monotonic tensor-to-scalar ratio: Natural Inflation} \label{NI}

The NI potential is given by
\begin{equation} 
V_{NI}=V_0\left[1+\cos\left(\frac{\phi}{f}\right)\right],
\label{vni}
\end{equation}
and from Eq.~(\ref{Slowpara}) it follows that the tensor-to-scalar ratio $r=16\epsilon$ is given by
\begin{equation} 
r=\frac{8}{f^2} \frac{\sin^2(\frac{\phi}{f})}{\left[1+\cos\left(\frac{\phi}{f}\right)\right]^2}.
\label{epsNI}
\end{equation}
quite clearly r grows without bound diverging for $\phi/f=\pi$, as illustrated in Fig.~\ref{FNI}. Typically inflation is terminated when $\epsilon=1.$ In NI  not only $\epsilon$ but also $\epsilon^{\prime} $ and $\epsilon^{\prime\prime } $ are monotonically increasing functions of $\phi$.
\begin{figure}[t!]
 \begin{center}
   \includegraphics[ width=8cm, height=6cm]{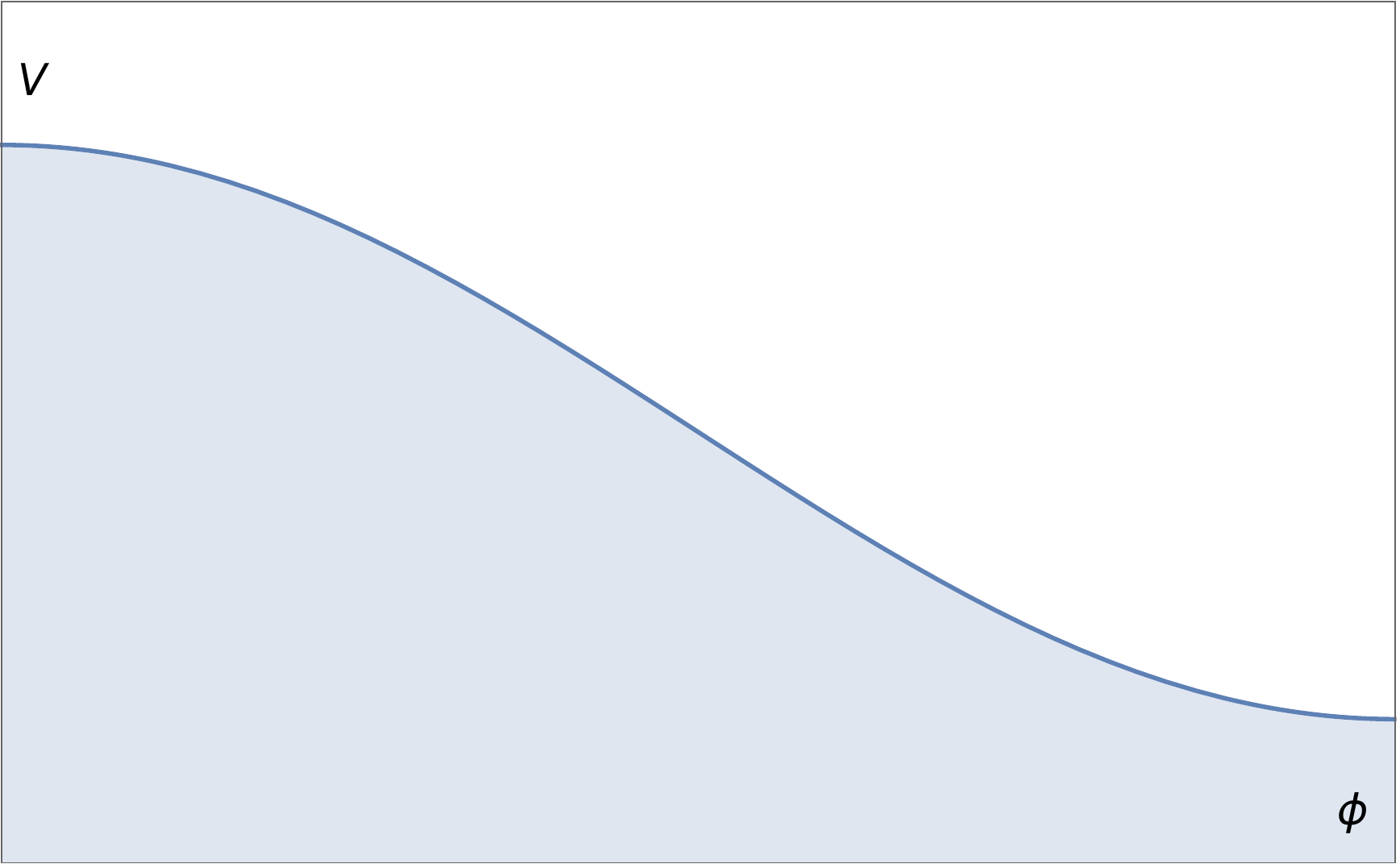}
    \includegraphics[ width=8cm, height=6cm]{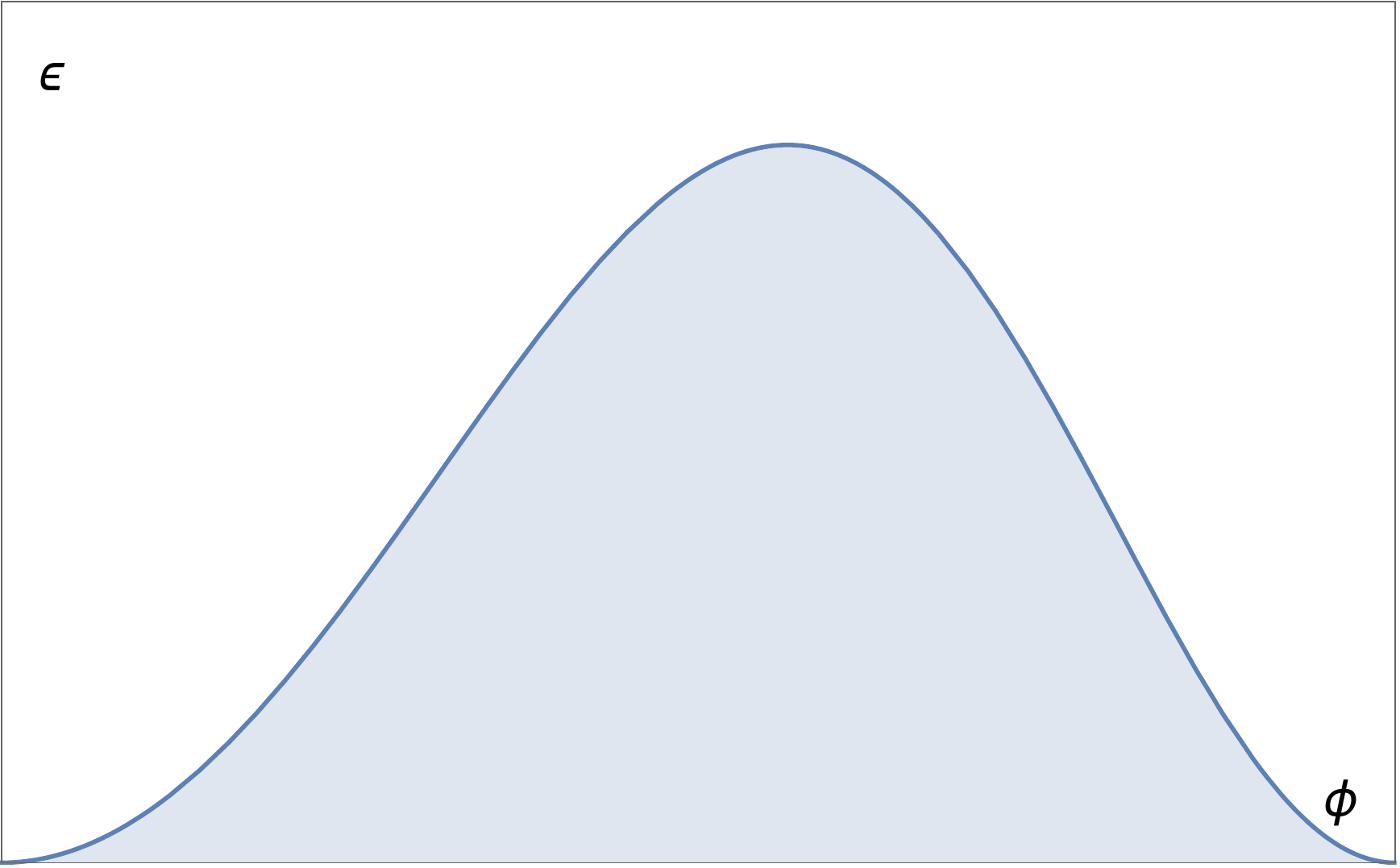}
        \caption{In Hybrid Natural Inflation there is a nearly flat plateau followed by a smooth fast transition to a lower flat constant plateau where most of inflation can occur and where inflation is terminated by a second waterfall-field. After the inflaton $\phi$ has carried out all of inflation the waterfall-field fast-roll towards the global minimum with vanishing energy (not shown). As a consequence of this the inflaton minimum has a non-vanishing energy (left) allowing for the possibility of a non-monotonic tensor-to-scalar ratio (right). }
\label{FHNI}
 \end{center}
\end{figure} 

\subsection{A model with a non-monotonic tensor-to-scalar ratio: Hybrid Natural Inflation} \label{HNI}
In HNI \cite{Ross:2009hg}-\cite{Ross:2016hyb} all of inflation is driven by the single-field $\phi$ although the end of inflation is triggered by a second, waterfall-field. The inflationary sector of HNI is 
\begin{equation} 
V_{HNI}=V_0\left[1+a\cos\left(\frac{\phi}{f}\right)\right],
\label{vhni}
\end{equation}
where $0\leq a<1$. The tensor-to-scalar ratio is given by
\begin{equation} 
r=\frac{8}{f^2}\frac{a^2 \sin^2(\frac{\phi}{f})}{\left[1+a\cos\left(\frac{\phi}{f}\right)\right]^2}.
\label{tensor}
\end{equation}
We see that $r=0$ at $\phi/f=0,\pi$ developing a maximum located at $\cos(\phi_{max}/f)=-a$ (see Fig.~\ref{FHNI}). The maximum value $r$ can take (away from $r_H$) is therefore 
$r_{max}=\frac{8 a^2 }{f^2 (1-a^2)}$. 
In both cases the running can be written as \cite{Ross:2016hyb}
\begin{equation} 
n_{sk}=\frac{r}{32}\left(3r-16\delta _{ns}+\frac{8}{f^2}              \right),
\label{running}
\end{equation}
without $a$ appearing explicitly. Thus, the running $n_{skH}$ will be negative if and only if
\begin{equation} 
f > \frac{1}{\sqrt{2\delta _{ns} -\frac{3}{8}r}}.
\label{scale}
\end{equation}
According to the values of Table 4 of Planck data \cite{Ade:2015lrj} we get $f>3.5$, in Planck units. In NI $f$ is strictly super-Planckian satisfying the previous bound with a negative running while in HNI $f$ can also be sub-Planckian with positive running allowing for the possibility of primordial black hole production during inflation   \cite{Kohri:2007qn}, \cite{Ross:2016hyb}.
%%%%%%%%%%%%%%%%%%%%%%%%%%%%%%%%%%%%%%%%%%%%%
%%%%%%%%%%%%%%%%%%%%%%%%%%%%%%%%%%%%%%%%%%%%%
\section{Conclusions}  \label{Conclusec}

We developed a model-independent study of single field slow-roll canonical inflation by imposing conditions on the slow-roll parameter $\epsilon(\phi)$ and its derivatives, $\epsilon^{\prime }(\phi)$ and $\epsilon^{\prime\prime }(\phi)$, to extract general 
conditions on the tensor-to-scalar ratio $r$ and the running $n_{sk}$. For models where $\epsilon(\phi)$ presents a maximum, a relatively large $r_H$ is most likely accompanied by a positive running, while a negligible tensor-to-scalar ratio typically implies negative running. The definitive answer, however, is given by the condition on the slow-roll parameter $\xi_2$, Eq.~(\ref{xi2}).
We have also shown that by imposing conditions  to the slow-roll parameter 
$\epsilon(\phi)$ and its derivatives $\epsilon^{\prime }(\phi)$ and $\epsilon^{\prime\prime }(\phi)$ we can accommodate sufficient inflation with a relatively large $r_H$ but still satisfying the
Planck \cite{Ade:2015lrj} or Planck-Keck-BICEP2 \cite{Ade:2015tva} constraints. The excursion of the field  $\Delta\phi_e\equiv |\phi_H-\phi_e | $ will be no larger than one if the function $\epsilon(\phi)$ has a maximum in a thin hill-shaped feature, decreasing to small values close to the end of 
inflation, at $\phi_e$. The maximum is required because observations indicate that $\epsilon_{H}$ is increasing in the observables scales, at 
$\phi_{H}$. Then $\epsilon(\phi)$ should decrease for $\phi>\phi_{max}$ for an effective way of generating the majority of $e$-folds of inflation. The contribution to the number of $e$-folds when $\epsilon{(\phi)}$ is growing can be 8 $e$-folds for $\Delta\phi_e$ less than one and $r_H$ close to the upper limit $r_H=0.12$. The end of inflation for vanishing $\epsilon(\phi)$ can be triggered by a hybrid field and a small $\Delta\phi_c\equiv |\phi_{max}-\phi_e | $ is obtained when $\epsilon(\phi)$ is sufficiently thin which, however, should not conflict with the other slow-roll parameter $\eta(\phi)$. Under these circumstances $\Delta\phi_e$ is restricted to a narrow windows of values.

\section{Acknowledgements}

G.G. gratefully acknowledges hospitality of the Rudolf Peierls Centre for Theoretical Physics, Oxford and useful discussions with Prof. Graham G. Ross 
and Shaun Hotchkiss. 
He also acknowledges financial support from PASPA-DGAPA, UNAM and CONACYT, Mexico. We all acknowledge support from 
\textit{Programa de Apoyo a Proyectos de Investigaci\'on e Innovaci\'on Tecnol\'ogica} (PAPIIT) UNAM, IN103413-3, \textit{Teor\'ias de Kaluza-Klein, 
inflaci\'on y perturbaciones gravitacionales}, IA101414, \textit{Fluctuaciones No-Lineales en Cosmolog\'ia Relativista} and IA103616, \textit{Observables en Cosmolog\'ia Relativista}. AHA is also grateful to SNI and PRODEP for partial financial support, and acknowledges a VIEP-BUAP-HEAA-EXC15-I research grant.


\begin{thebibliography}{99}  
\bibitem{Guth:1981} A.H.~Guth, 
{\it The Inflationary Universe: A possible solution to the horizon and flatness problems} 
Phys.\ Rev.\ \textbf{D23} (1981) 347. %%CITATION = PRLTA,65,3233;%%

\bibitem{Linde:1982} A.D~Linde, 
{\it A New Inflationary Universe Scenario: A Possible Solution of the Horizon, Flatness, Homogeneity, Isotropy and Primordial Monopole Problems} 
Phys.\ Lett.\ \textbf{B108} (1982) 389. %%CITATION = PRLTA,65,3233;%%

\bibitem{Albrecht:1982} A.~Albrecht and P.J.~Steinhardt, 
{\it Cosmology for Grand Unified Theories with Radiatively Induced Symmetry Breaking}
Phys.\ Rev.\ Lett.\ \textbf{48} (1982) 1220. %%CITATION = PRLTA,65,3233;%%

%\cite{Lyth:1998xn}
\bibitem{Lyth:1998xn} For reviews see e.g., 
  D.~H.~Lyth and A.~Riotto,
{\it Particle physics models of inflation and the cosmological density perturbation,}
  Phys.\ Rept.\  {\bf 314} (1999) 1
  [hep-ph/9807278];
  %%CITATION = HEP-PH/9807278;%%
  %1242 citations counted in INSPIRE as of 06 Oct 2014;
  D.~H.~Lyth and A.~R.~Liddle,
 {\it The primordial density perturbation: Cosmology, inflation and the origin of structure,}
  Cambridge, UK: Cambridge Univ. Pr. (2009) 497 p;
  %17 citations counted in INSPIRE as of 06 Oct 2014;
D.~Baumann,
  {\it TASI Lectures on Inflation,}
  arXiv:0907.5424 [hep-th].
  %%CITATION = ARXIV:0907.5424;%%
  %194 citations counted in INSPIRE as of 06 Oct 2014


%\cite{Hu:2011vr}
\bibitem{Hu:2011vr}
  W.~Hu,
  {\it Generalized Slow Roll for Non-Canonical Kinetic Terms,}
  Phys.\ Rev.\ D {\bf 84} (2011) 027303
  [arXiv:1104.4500 [astro-ph.CO]].
  %%CITATION = ARXIV:1104.4500;%%
  %26 citations counted in INSPIRE as of 30 Jul 2014
  
    %\cite{Ade:2015lrj}
\bibitem{Ade:2015lrj}
  P.~A.~R.~Ade {\it et al.} [Planck Collaboration],
  {\it Planck 2015 results. XX. Constraints on inflation,}
  arXiv:1502.02114 [astro-ph.CO].
  %%CITATION = ARXIV:1502.02114;%%
  %240 citations counted in INSPIRE as of 20 ao�t 2015
      
  
  %\cite{Ade:2015tva}
\bibitem{Ade:2015tva}
  P.~A.~R.~Ade {\it et al.} [BICEP2 and Planck Collaborations],
  {\it Joint Analysis of \textrm{BICEP2/Keck Array} and \textrm{Planck} Data,}
  Phys.\ Rev.\ Lett.\  {\bf 114} (2015) 101301
  [arXiv:1502.00612 [astro-ph.CO]].
  %%CITATION = ARXIV:1502.00612;%%
  %176 citations counted in INSPIRE as of 20 ao�t 2015                                             

%\cite{Lyth:1997}
\bibitem{Lyth:1997}
D.~H.~Lyth,
{\it What would we learn by detecting a gravitational wave signal in the cosmic microwave background anisotropy?}
Phys. Rev. Lett. 78(1997)1861 [hep-ph/0502047].
  %%CITATION = HEP-PH/0502047;%%
  %138 citations counted in INSPIRE as of 08 May 2014

%\cite{Boubekeur:2005zm}
\bibitem{Boubekeur:2005zm}
  L.~Boubekeur and D.~H.~Lyth,
  {\it Hilltop inflation,}
  JCAP {\bf 0507} (2005) 010
  [hep-ph/0502047].
  %%CITATION = HEP-PH/0502047;%%
  %138 citations counted in INSPIRE as of 08 May 2014

%\cite{BenDayan:2009kv}
\bibitem{BenDayan:2009kv}
  I.~Ben-Dayan and R.~Brustein,
  ``Cosmic Microwave Background Observables of Small Field Models of Inflation,''
  JCAP {\bf 1009} (2010) 007
  [arXiv:0907.2384 [astro-ph.CO]].
  %%CITATION = ARXIV:0907.2384;%%
  %35 citations counted in INSPIRE as of 22 May 2014

%\cite{Hotchkiss:2011gz}
\bibitem{Hotchkiss:2011gz}
  S.~Hotchkiss, A.~Mazumdar and S.~Nadathur,
  ``Observable gravitational waves from inflation with small field excursions,''
  JCAP {\bf 1202} (2012) 008
  [arXiv:1110.5389 [astro-ph.CO]].
  %%CITATION = ARXIV:1110.5389;%%
  %46 citations counted in INSPIRE as of 22 May 2014

%\cite{Antusch:2014cpa}
\bibitem{Antusch:2014cpa}
  S.~Antusch and D.~Nolde,
  {\it BICEP2 implications for single-field slow-roll inflation revisited,}
  arXiv:1404.1821 [hep-ph].
  %%CITATION = ARXIV:1404.1821;%%
  %12 citations counted in INSPIRE as of 22 May 2014
  
  %\cite{Chatterjee:2014hna}
\bibitem{Chatterjee:2014hna}
  A.~Chatterjee and A.~Mazumdar,
 {\it Bound on largest $r\lesssim 0.1$ from sub-Planckian excursions of inflaton,}
  JCAP {\bf 1501} (2015) 01,  031
  doi:10.1088/1475-7516/2015/01/031
  [arXiv:1409.4442 [astro-ph.CO]].
  %%CITATION = doi:10.1088/1475-7516/2015/01/031;%%
  %5 citations counted in INSPIRE as of 05 Feb 2016
    
%\cite{Kohri:2007qn}
\bibitem{Kohri:2007qn}
  K.~Kohri, D.~H.~Lyth and A.~Melchiorri,
 {\it Black hole formation and slow-roll inflation,}
  JCAP {\bf 0804} (2008) 038
  [arXiv:0711.5006 [hep-ph]].
  %%CITATION = ARXIV:0711.5006;%%
  %50 citations counted in INSPIRE as of 06 Nov 2015

%\cite{Vazquez:2014uca}
\bibitem{Vazquez:2014uca}
  J.~A.~Vazquez, M.~Carrillo-Gonzalez, G.~German, A.~Herrera-Aguilar and J.~C.~Hidalgo,
  ``Constraining Hybrid Natural Inflation with recent CMB data,''
JCAP {\bf 1502} (2015) 02,  039
[arXiv:1411.6616 [astro-ph.CO]].
%%CITATION = ARXIV:1411.6616;%%
  %1 citations counted in INSPIRE as of 29 Sep 2015

\bibitem{Liddle:2000cg} 
A.~R.~Liddle and D.~H.~Lyth, 
\textit{Cosmological Inflation and Large-Scale Structure}, 
%``Cosmological inflation and large-scale structure,''
%\href{http://www.slac.stanford.edu/spires/find/hep/www?irn=4458796}{SPIRES entry}
Cambridge University Press, (2000).
  
\bibitem{Freese:1990rb} K.~Freese, J.~A.~Frieman and A.~V.~Olinto, 
{\it Natural inflation with pseudo - Nambu-Goldstone bosons},
Phys.\ Rev.\ Lett.\ \textbf{65} (1990) 3233. %%CITATION = PRLTA,65,3233;%%

\bibitem{Freese:1993} F.~Adams, J.R.~Bond, K.~Freese, J.~A.~Frieman
  and A.~V.~Olinto, 
{\it Natural inflation: Particle physics models, power law spectra for large scale structure, and constraints from COBE}
Phys.\ Rev.\textbf{D47} (1993) 426-455. %%CITATION = PRLTA,65,3233;%%

\bibitem{Freese:2008if} K.~Freese, C.~Savage and W.~H.~Kinney, 
{\it Natural Inflation: the status after WMAP 3-year data},
Int.\ J.\ Mod.\ Phys.\ D \textbf{16} (2008) 2573 [arXiv:0802.0227 [hep-ph]]. 
%%CITATION = IMPAE,D16,2573;%%
%\cite{Freese:2014nla}

%\cite{Freese:2014nla}
\bibitem{Freese:2014nla}
  K.~Freese and W.~H.~Kinney,
  {\it Natural Inflation: Consistency with Cosmic Microwave Background Observations of Planck and BICEP2,}
  arXiv:1403.5277 [astro-ph.CO].
  %%CITATION = ARXIV:1403.5277;%%
  %46 citations counted in INSPIRE as of 06 Oct 2014
  
  %\cite{Ross:2009hg}
  \bibitem{Ross:2009hg}
  G.~G.~Ross and G.~German,
  {\it Hybrid natural inflation from non Abelian discrete symmetry,}
  Phys.\ Lett.\ B {\bf 684} (2010) 199
  [arXiv:0902.4676 [hep-ph]].
  %%CITATION = ARXIV:0902.4676;%%
  %11 citations counted in INSPIRE as of 08 Oct 2014

%\cite{Ross:2010fg}
\bibitem{Ross:2010fg}
  G.~G.~Ross and G.~German,
  {\it Hybrid Natural Low Scale Inflation,}
  Phys.\ Lett.\ B {\bf 691} (2010) 117
  [arXiv:1002.0029 [hep-ph]].
  %%CITATION = ARXIV:1002.0029;%%
  %8 citations counted in INSPIRE as of 08 Oct 2014

%\cite{Carrillo-Gonzalez:2014tia}
\bibitem{Carrillo-Gonzalez:2014tia}
  M.~Carrillo-Gonzalez, G.~German, A.~Herrera-Aguilar, J.~C.~Hidalgo and R.~Sussman,
  {\it Testing Hybrid Natural Inflation with BICEP2,}
Phys.\ Lett.\ B {\bf 734} (2014) 345
[arXiv:1404.1122 [astro-ph.CO]].
%%CITATION = ARXIV:1404.1122;%%
  %7 citations counted in INSPIRE as of 29 Sep 2015
  
  %\cite{Ross:2016hyb}
\bibitem{Ross:2016hyb}
  G.~G.~Ross, G.~German and J.~A.~Vazquez,
   {\it Hybrid Natural Inflation,}
  %``Hybrid Natural Inflation,''
  arXiv:1601.03221 [astro-ph.CO].
  %%CITATION = ARXIV:1601.03221;%%
  
\end{thebibliography}
\end{document}